\documentstyle[aps,prb,psfig]{revtex}
\title{Monte-Carlo Simulations of the Dynamical Behavior of the
Coulomb Glass}
\author{Torsten Wappler \dag, Thomas Vojta \dag\ddag\ and Michael
Schreiber
 \dag\\ \small\dag\ Institut f\"ur Physik, Technische
Universit\"at,\\
\small D-09107 Chemnitz, Federal Republic of Germany\\
\small \ddag\ Materials Science Institute, University of Oregon,\\
\small Eugene, OR 97403, USA}
\def\halb{\raisebox{0.3ex}{1}\!/\!\raisebox{-0.8ex}{2}}
\let\half=\halb
\begin{document}
\maketitle
\begin{abstract}
We study the dynamical behavior of disordered many-particle systems
with
long-range Coulomb interactions by means of damage-spreading
simulations.
In this type of Monte-Carlo simulations one investigates the time
evolution of the damage, i.e. the difference of the occupation
numbers
of two systems, subjected to the same thermal noise. We analyze the
dependence
of the damage on temperature and disorder strength. For zero disorder
the spreading transition coincides with the equilibrium phase
transition,
whereas for finite disorder, we find evidence for a
dynamical phase transition well below the transition temperature of
the
pure system.
\end{abstract}

\section{Introduction}
The combined influence of disorder and long-range interactions on the
properties of many-particle systems has been a subject of great
interest
for some time. In electronic systems already disorder or interactions
alone
can drastically change the physical behavior. Disorder can lead,
e.g., to a
metal-insulator transition due to Anderson localization. On the other
hand,
a metal-insulator transition can also be induced by correlations due
to
electron-electron interactions. If disorder and interactions are both
significant then complex physical problems and phenomena arise, many
of which
are not completely understood.

The behavior of strongly localized correlated electrons in disordered
insulators is especially complicated, both experimentally and
theoretically.
Thus progress has been slow since the first
investigations.\cite{pollak,efros}
Many properties of such systems are still poorly understood. In
particular
there are only few and contradicting results on thermodynamics, phase
diagram,
phase transitions or critical behavior, and the examination of the
dynamical
behavior is only at its beginning.\cite{pollak92}
Two of the central questions are whether or not the disordered
interacting
electron system shows glassy behavior and what is the nature of the
glassy
"state". Two different views can be found in the literature. In the
earlier
work the formal similarity between disordered localized electrons and
spin glasses had lead to speculations about a possible equilibrium
phase
transition to a spin-glass-like low-temperature
phase.\cite{davies,gruene}
More recent investigations show, however, growing experimental and
theoretical
evidence of the transition being of dynamical nature.
\cite{benchorin,baranovskii,vojta93,tenelsen}

In this paper we study the dynamical behavior of disordered localized
electrons by means of the damage-spreading method. In this type of
Monte-Carlo simulations the microscopic differences of the time
evolution
between two systems are investigated.
In particular, we address the question
of a dynamical phase transitions from a
dynamically active high-temperature phase to a frozen low-temperature
phase
upon changing characteristic parameters like disorder or temperature.
Our paper is organized as follows. In Sect. \ref{sec:model} we
introduce the
Coulomb glass model, the prototype model of disordered localized
electrons.
In Sect. \ref{sec:ds} we describe the damage spreading technique,
whereas in
Sect. \ref{sec:res} we present the results for the dynamical behavior
of the
model. Section \ref{sec:concl} is dedicated to some discussions and
conclusions.

\section{Model}
\label{sec:model}
Our investigations are based on the Coulomb glass model first
proposed by Efros
and Shklovskii \cite{efros} to describe compensated doped
semiconductors.
Later it has also been applied to simulate granular metals
\cite{adkins} and
conducting polymers.\cite{phillips,schreiber96}
The model consists of a square or cubic lattice of linear size $L$
with $N=L^d$
sites (in $d$ dimensions) and lattice constant $a$. The sites can be
occupied
by
$KN$ ($0<K<1$) electrons. These electrons are interacting via an
unscreened
Coulomb potential. To guarantee charge neutrality every site carries
a
compensating charge of $+Ke$ ($-e$ is the charge of the electron).
The disorder
of this system is described by the random potential $\varphi_i$. The
Hamiltonian of the Coulomb glass is given by
\begin{equation}
H = \sum_{i} (\varphi_i-\mu)n_i +
\frac{1}{2}\sum_{i\not=j}(n_i-K)(n_j-K)
U_{ij}\qquad U_{ij} = \frac{e^2}{r_{ij}}
\label{cg_hamil}
\end{equation}
where $\mu$ is the chemical potential, $n_i$ (with values 0 or 1) is
the occupation number of site $i$ and $r_{ij}$ denotes the distance
between
sites $i$ and $j$. In the rest of the paper we set the interaction
strength
between nearest neighbor sites $e^2/a=1$ which fixes the energy
scale.
The random potential energies $\varphi_i$ are independent from
each other and chosen according to some probability distribution
$W(\varphi)$.
We use the box distribution with mean 0 and width $W_0$. The
parameter $W_0$
measures the strength of the disorder. Specifically, we investigate a
half-filled system ($K=\halb$). Then the Coulomb glass model is
particle-hole
symmetric and the chemical potential vanishes. (Note that the two
quantities
$K$ and $\mu$ are not independent of each other. We treat $K$ as a
free
parameter and calculate $\mu$ from it.)

For later reference we briefly mention some properties of the Coulomb
glass
model. One of the central quantities is the
single-electron density of states
\begin{equation}
  g(\varepsilon, T)=\frac{1}{N}\sum_i
\langle\delta(\varepsilon-\varepsilon_i)
\rangle
\end{equation}
at energy $\varepsilon$ and temperature $T$, where
$\langle\ldots\rangle$
denotes thermal and disorder averages. $\varepsilon_i$ are the
single-electron
energies given by
\begin{equation}
  \varepsilon_i = \varphi_i -\mu + \sum_{j\neq i} U_{ij}(n_j-K).
\end{equation}

\begin{figure}[htbp]
  \centerline{\psfig{figure=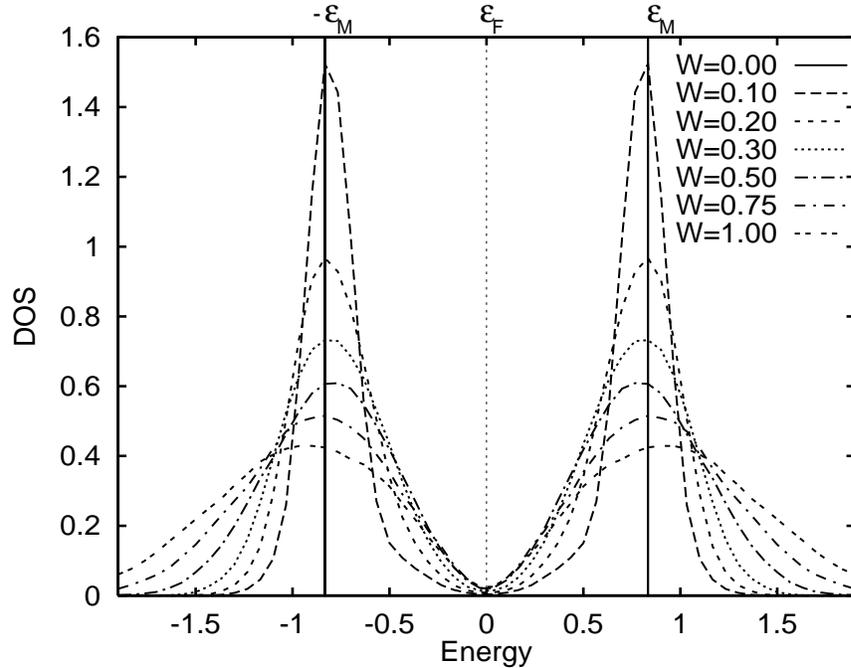}}
  \caption{Single-electron density of states of the Coulomb glass at
$T=0.008$
for different strengths of disorder. $\varepsilon_M$ indicates the
Madelung
energy, $\varepsilon_F$ the Fermi energy.}
  \label{fig:dos}
\end{figure}

The single-electron density of states of the Coulomb glass shows a
pronounced
gap, called the Coulomb gap, close to the Fermi energy
$\varepsilon_F$
(see Fig. \ref{fig:dos}). At zero temperature the density of states
actually
vanishes at the Fermi energy \cite{efros}, close to the Fermi energy
it can
be described by a power law
\begin{equation}
g(\varepsilon) \propto \vert \varepsilon - \varepsilon_F \vert^\alpha
\label{eq:gap}
\end{equation}
where $\alpha$ is approximately 1.2 for two-dimensional (2D) and 2.5
for 3D
systems.\cite{moebius}
At finite temperatures the Coulomb gap is filled gradually (for
recent
simulation results see, e.g., Ref. \cite{sarv}).

The Coulomb glass model (\ref{cg_hamil}) describes a system without
internal
dynamics. In reality the electrons, though localized, are coupled to
additional
(vibrational) degrees of freedom, which lead to transitions between
the
many-electron states. Phenomenologically this can be
simulated by a Monte-Carlo method. In every Monte-Carlo step we
change
the occupation numbers of one or several sites with a certain
probability.
Within the Metropolis algorithm this probability is given by
\begin{equation}
P = \left\{
\begin{array}{lr}
1 & , \Delta H < 0 \\
\exp[-\frac{\Delta H}{k_{B}T}] & , \Delta H > 0 \\
\end{array}
\right.
\label{trans_p}
\end{equation}
where $\Delta H$ is the energy difference between the many-particle
states
before and after such a change, and $k_B$ is the Boltzmann constant.
$N$ such Monte-Carlo steps are called a Monte-Carlo sweep
which is the natural time scale of our calculations.

To simulate the dynamics one can use different ``move classes'',
which determine how the occupation numbers are changed in every
Monte-Carlo
step to get the new configuration. The simplest move class
consists of exchanging a single electron with a
reservoir (i.e. the conduction band in the case of doped
semiconductors),
other classes include hopping of single electrons between the sites,
or correlated hopping of several
electrons. In this paper we present results obtained by using only
single-electron exchanges between the system and a
reservoir, but we have also checked more complicated move classes.
As long as we do not include distance-dependent ''tunneling terms''
into the transition probabilities (\ref{trans_p}), applying different
move
classes yields data which do not show a qualitatively different
behavior.
We attribute this result to the fact that single and multiple
electron hops
can be combined from the moves in our implementation of single
electron
exchanges with an external reservoir. Thus all many-electron states
with
$KN$ electrons are available in our simulation. A more detailed
investigation
of
this question including the effects of distance-dependent transition
probabilities on the damage-spreading simulations is in progress.

\section{Damage Spreading}
\label{sec:ds}
The damage-spreading technique \cite{stauffer} is a modification of
the usual
monte-Carlo method. The idea is to look not at the time evolution of
a single system but to compare the time evolutions of {\em two}
systems which
are subjected to the same thermal noise (i.e., the same random
numbers are used
within the Metropolis algorithm). Usually,
at the beginning of the simulation the occupation numbers of both
systems differ only at a single site (or at
a few sites, e.g. a single column in a 2D lattice system).

Since both systems are thermodynamically identical, averages of
equilibrium
quantities will be the same for both systems. Microscopically,
however, the
two systems may evolve differently from each other.
The central observable in damage-spreading simulations is the
Hamming distance $D(t)$, which is the portion of sites for which the
occupation numbers differ between the two systems. $D(t)$, which
measures
the  "damage", is given by
\begin{equation}
D(t) = \frac{1}{N}\sum_{i}\vert n_i^{\rm o}(t)-n_i^{\rm c}(t)\vert
\end{equation}
where $n_i^{\rm o}(t)$and  $n_i^{\rm c}(t)$ are the occupation
numbers of site
$i$ of the original system and the copy at (Monte-Carlo) time $t$.
For $D(t)=0$ the two systems are identical, $D(t)=\halb$
describes completely uncorrelated
configurations, and for $D(t)=1$ the two systems are totally
anticorrelated.
In the course of the time evolution the two systems evolve towards a
steady
state, in which $D(t)$ fluctuates around an asymptotic average value
\begin{equation}
D= \lim_{\tau \to \infty} \lim_{t \to \infty}
 \frac 1 \tau \int_t^{t+\tau}  d t' D(t')
\end{equation}

Depending on the values of the external parameters temperature and
disorder
different regimes can be observed in principle if the initial damage
$D(0)$ is small: The damage may heal out during the time evolution
($D=0$), the systems may stay partially correlated for infinite
time ($D<\halb$), or the systems may
become completely uncorrelated so that $D=\halb$. In contrast to
the thermodynamics the detailed behavior of $D(t)$ depends on
the choice of the dynamical algorithm. Whereas Metropolis, Glauber
and
heat-bath dynamics give the same results for equilibrium quantities
of
a single system, the damage spreading results differ.
For the Metropolis dynamics which we use (as
well as for the Glauber dynamics) the damage tends to heal at
low temperatures and tends to spread at high temperatures.
\cite{stauffer} In contrast, the heat-bath dynamics yields
healing at high temperatures and frozen configurations
at low temperatures.\cite{derrida}
(Note that since $D$ is {\em not} a thermodynamic quantity but
measures
the microscopic differences between two systems, there is no reason
to expect that different dynamical algorithms give the same results.)

We apply the damage-spreading technique to the 2D Coulomb-glass
model at half filling $K=\halb$ and linear system sizes
$L=20,...,80$.
The simulation proceeds as follows: (i) We create the initial system
by
choosing random potential values according to the probability
distribution $W(\varphi)$ and occupy the sites at random
with $KN$ electrons. (ii) We equilibrate this system at
temperature $T$ by
performing several (at least 300) Monte-Carlo sweeps according to the
Metropolis algorithm. (iii) A copy of the system is created and
modified at a
single site (or several sites). This difference in the occupation
numbers
constitutes the initial damage. (iv) We study the time evolution of
the
original and the copy using the same random numbers in the Metropolis
algorithm
for both systems. The damage $D(t)$ is recorded and its asymptotic
value $D$
is determined.

Note that there is a modification of the damage-spreading method that
can be
used to determine {\em equilibrium} quantities instead of purely
dynamic ones.
\cite{coniglio,glotzer} In that kind of simulations the occupation
number
of a single site in one of the systems is fixed whereas it is allowed
to
fluctuate in the other system. Consequently, the two systems are
thermodynamically {\em different} and the damage can be related to
equilibrium
correlation functions. Since in this paper we are interested in the
properties
of the dynamics rather than in equilibrium quantities, our data is
gained by means of the original damage-spreading method, where the
occupation numbers of {\em both} systems are allowed to fluctuate.

\section{Results}
\label{sec:res}
\subsection{Time evolution}
\begin{figure}[htbp]
  \centerline{\psfig{figure=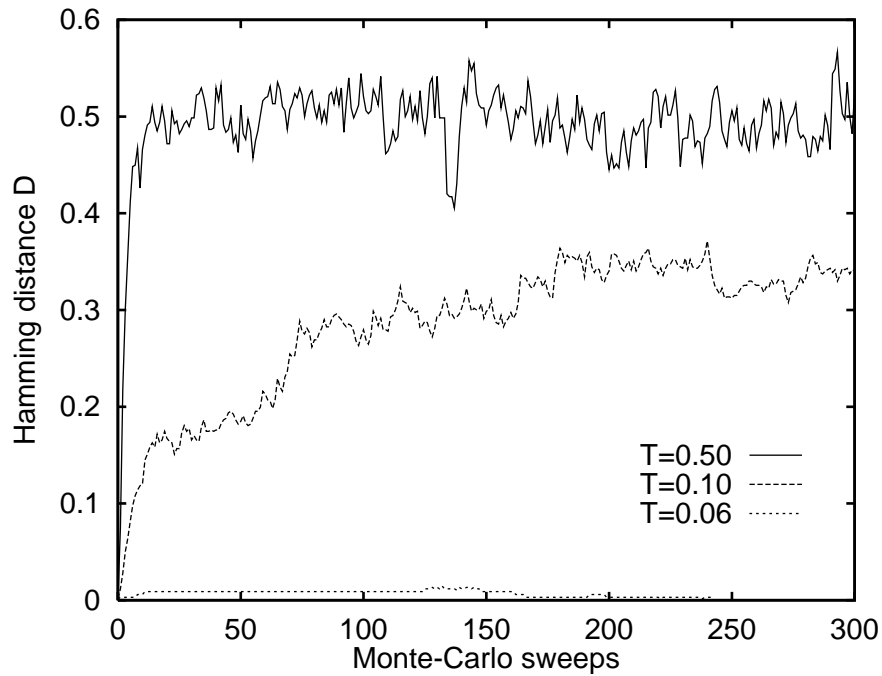}}
  \caption{Time dependence of the Hamming distance of the 2D Coulomb
glass for
different temperatures and $W_0=0$.}
  \label{fig:d_time00}
\end{figure}

\begin{figure}[htbp]
  \centerline{\psfig{figure=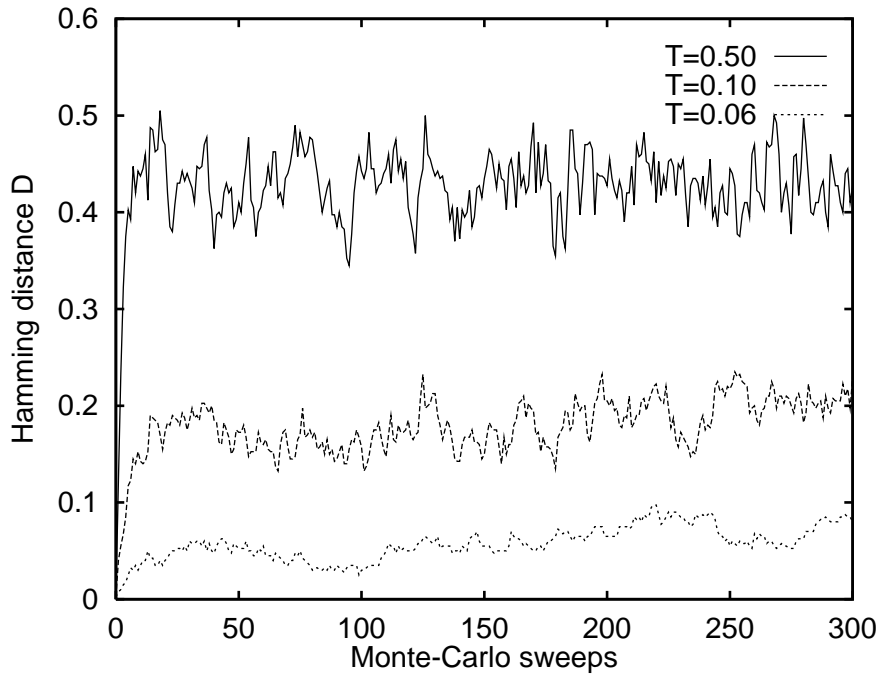}}
  \caption{Time dependence of the Hamming distance of the 2D Coulomb
glass for
different temperatures and $W_0=0.5$.}
  \label{fig:d_time05}
\end{figure}

In this subsection we present data on the time evolution of the
damage $D(t)$ starting with an initial damage consisting of a single
site.
In analogy to the well studied 2D Ising model
\cite{stauffer,stauffer2,campbell} we find
that for temperatures below a certain temperature $T_{\rm s}$, called
the
spreading temperature, the damage $D(t)$ remains small and
eventually heals, giving an asymptotic value of $D=0$.
For temperatures larger than $T_{\rm s}$ the damage increases with
time
until a steady state is reached where $D(t)$ fluctuates around a
finite value.
Consequently, the asymptotic damage $D$ is finite in this regime.
In Fig. \ref{fig:d_time00} the time evolution
of $D(t)$ is shown for the Coulomb glass with zero disorder $W_0=0$.
The three curves presented correspond to the three regimes
discussed in the last section. At $T=0.5$ the damage increases
quickly and
then fluctuates around $D=\halb$. This means the two
systems become completely uncorrelated very fast. Consequently
we are above the spreading temperature $T_{\rm s}$. At $T=0.1$ the
evolution
of $D(t)$ is much
slower and the asymptotic damage is smaller than $\halb$. This
behavior
occurs, because the system is in the vicinity of the spreading
transition at
$T_{\rm s}$. It corresponds to the critical slowing down in
ordinary critical phenomena.
At $T=0.06$ the damage remains small and eventually heals, thus the
system is below the spreading temperature $T_{\rm s}$.
In the case of finite disorder $W_0$ the time evolution of the damage
is similar (see Fig. \ref{fig:d_time05}).
The asymptotic damage $D$ is, however, different from 0 or $\half$
even far
away
from the spreading transition.
The dependence of the damage on the external parameters temperature
and
disorder
 is investigated in more detail in Subsect. \ref{sec:asymp}.

\subsection{Influence of the long-range interaction}

\begin{figure}[htbp]
  \centerline{\psfig{figure=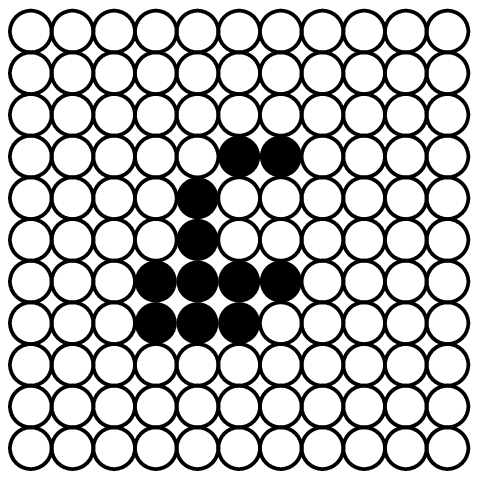,width=6cm}\hspace{1cm}
\psfig{figure=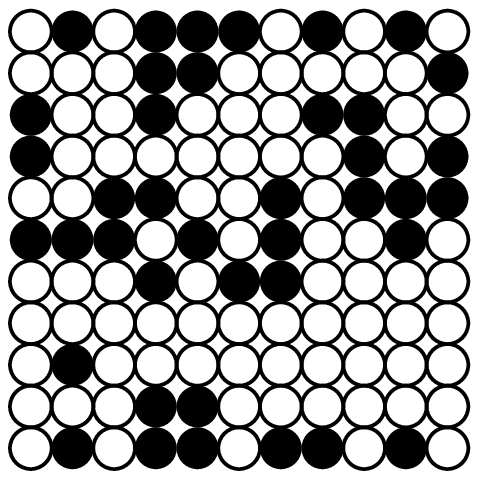,width=6cm}}
  \caption{Snapshot of the damage for 2D systems with short-range
interactions
   ({\em left}) and long-range interactions ({\em right}) for
$T=0.5$ and
   $W_0=0.5$ at a time of $5$ Monte-Carlo sweeps after the
introduction
   of a single damaged site. A filled circle indicates a damaged
site where
the occupation
   numbers of the two systems differ, an empty circle indicates that
the
   occupation numbers of that site are identical in both systems.}
  \label{fig:kull}
\end{figure}

The character of the interaction has a large influence on the time
evolution of the damage. In systems with nearest-neighbor
interactions,
e.g. the Ising model, the damage can only spread within a single
Monte-Carlo
step from one site of the system to its neighbor.
Therefore the clouds of damaged sites can grow only slowly in
space and tend to be more compact (but not necessarily
connected). In contrast, in systems with long-range
interactions the occupation number of any  site  effects {\em all}
other sites. The damage can spread from one site of the system
to any other site within a single Monte-Carlo step. Therefore the
damage
spreads much faster as in systems with short-range interactions
and the damage clouds are usually not compact. A comparison of the
two cases is presented in Fig. \ref{fig:kull}.

Note, that since the damage can spread from one site to any other
site in
the case of long-range interactions, some of the methods developed to
analyze the damage-spreading simulations \cite{stauffer,stauffer2}
cannot be
used for systems with long-range interactions. This applies to
all methods that measure the spatial extension of the damage
and its evolution, because the spatial extent of the damage cloud
is not a well defined quantity for systems with
long-range interactions.

\subsection{Temperature and disorder dependence of the asymptotic
damage}
\label{sec:asymp}
\begin{figure}[htbp]
  \centerline{\psfig{figure=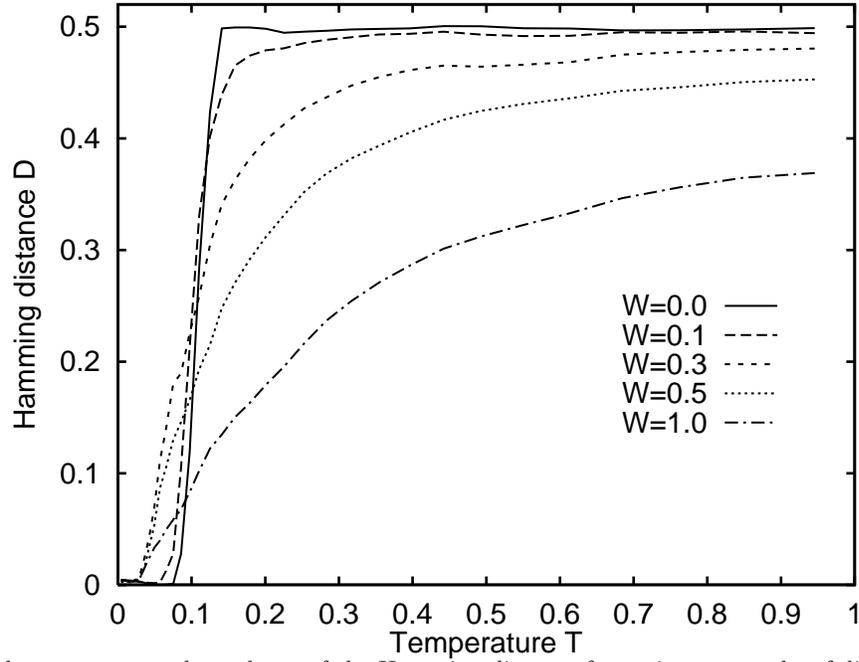}}
  \caption{Overview of the temperature dependence of the Hamming
distance for
various strengths of disorder of a 2D system with $N=20^2$ sites.}
  \label{fig:over}
\end{figure}

\begin{figure}[htbp]
  \centerline{\psfig{figure=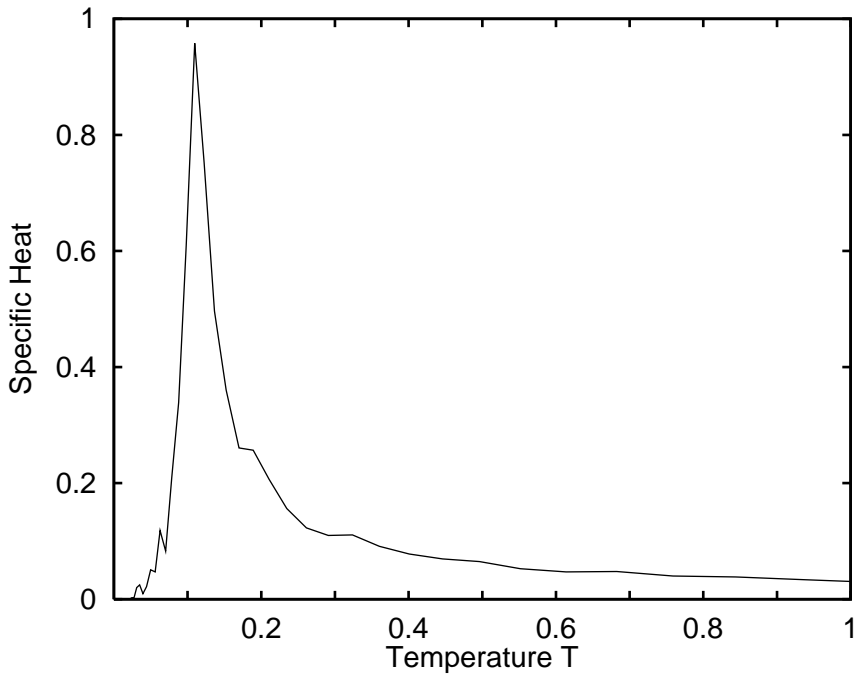}}
  \caption{Specific heat of the 2D Coulomb glass at $W_0=0$
calculated via the
derivation of the internal energy of the system.}
  \label{fig:c_v}
\end{figure}

We now turn to the main results of this paper. Figure \ref{fig:over}
shows an
overview of the temperature and disorder dependence of the asymptotic
Hamming distance  $D$.
For disorder strength $W_0=0$ there is a pronounced transition at
a spreading temperature of approximately $T_{\rm s}=0.1$ between a
low-temperature regime with $D=0$ and a high-temperature regime with
$D=\halb$.
Within our numerical accuracy the spreading temperature $T_{\rm s}$
coincides
with the equilibrium critical point $T_{\rm c}$ of the model without
disorder
which we determined from the peak in the specific heat $C_{\rm v}$ of
the Coulomb glass model as a function of temperature
(see Fig. \ref{fig:c_v}).
For very high temperatures $T\rightarrow\infty$ the spreading of the
damage
is drastically slowed down due to the fact that the probability $P$
in the
Metropolis algorithm, Eq. \ref{trans_p}, becomes independent of the
actual configurations of the two systems (original and copy)
and reaches  $P=1$. This means
that in both systems nearly every exchange of electrons is performed
and
differences in the occupation numbers occur only rarely.
Our investigations of the spreading behavior for very high
temperatures
show that the Hamming distance $D$ still reaches a plateau if plotted
versus
time as in Figs. \ref{fig:d_time00} and \ref{fig:d_time05},
but the relaxation time diverges as is predicted in a recent
mean-field theory.\cite{meanfield}
The damage-spreading transition in the Coulomb glass model without
disorder
occurs thus in complete analogy to that in the Ising
model.\cite{stauffer}

For finite disorder strength $W_0$, however, this behavior changes in
several aspects. First, the values of the asymptotic Hamming distance
in the high-temperature regime are smaller than $D=\halb$. This
means, the two
systems remain partially correlated even for high temperatures. The
reason for
that is easy to understand: In the presence of a random potential the
electrons
are trapped (repulsed) at sites with small (high) potential
values $\varphi_i$. These sites are identical in the original system
and its copy. Therefore the presence of a random potential tends to
reduce the damage. With increasing strength of disorder this trapping
effect becomes larger, so that the maximum value of the damage is
more and more reduced. On the other hand, increasing temperature
makes it
easier
to overcome the potential differences so that the described reduction
of the
damage becomes less effective.

\begin{figure}[htbp]
  \centerline{\psfig{figure=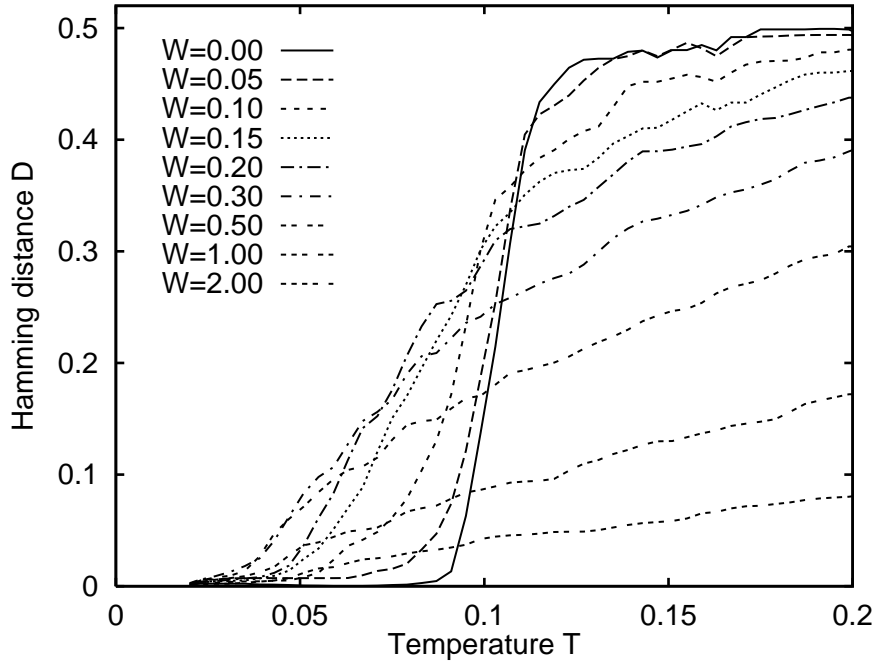}}
  \caption{Hamming distance versus temperature for various strengths
of
disorder of a 2D system with $N=20^2$ sites.}
  \label{fig:diff_w}
\end{figure}

The second effect of the disorder concerns the behavior of $D$ at
low temperatures
and close to the spreading point. This region is shown in more detail
in
Fig. \ref{fig:diff_w}.
In the case of finite disorder the asymptotic damage remains
finite even at temperatures below the spreading temperature of the
model
without
disorder. This somewhat counterintuitive result, viz. an acceleration
of the
dynamics by disorder, can be understood by looking at the
single-electron
density of states of the Coulomb glass model (see Fig.
\ref{fig:dos}).
For $W_0=0$ the single-electron density of states at low temperatures
has a
hard gap around the Fermi energy $\varepsilon_F=0$ and two peaks
at the Madelung energies $\pm\varepsilon_M$. Therefore there are only
exponentially few sites that can be excited at low temperatures and
thus the Hamming distance vanishes.
In contrast, for finite disorder $W_0$, the gap in the density of
states
is not exponential but the power-law Coulomb gap (\ref{eq:gap}).
Therefore more sites can be excited at low temperatures and the
dynamics
does not freeze completely, i.e., the Hamming distance remains
finite.

As can be seen in Fig. \ref{fig:diff_w}, even for finite disorder
strength
$W_0$ there is, however, a spreading temperature $T_{\rm s}(W_0)$,
below which
the asymptotic damage vanishes. $T_{\rm s}(W_0)$ decreases with
increasing
$W_0$, but seems to tend to a finite limiting value for large $W_0$
which
we approximately determined to $T_{\rm s} (\infty)\approx0.03$.
Note, that the existence of a spreading transition in the case of
finite
disorder is a purely dynamic phenomenon, since the system does not
undergo
an equilibrium phase transition.
\begin{figure}[htbp]
  \centerline{\psfig{figure=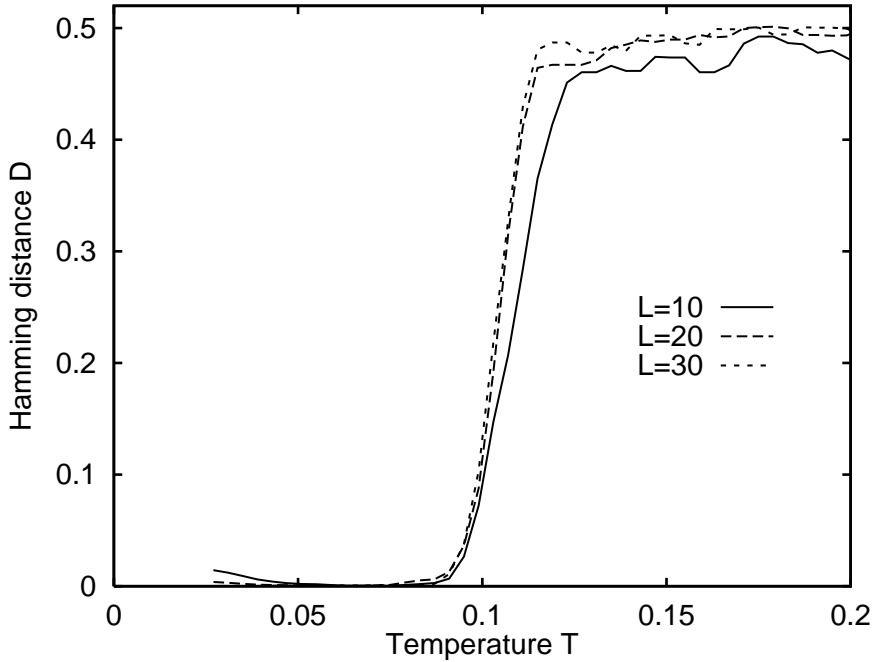}}
  \caption{Hamming distance versus temperature for various system
sizes at
$W_0=0$.}
\label{fig:finit}
\end{figure}

In order to determine more detailed properties of the spreading
transition
a careful analysis of finite size effects is necessary. In Fig.
\ref{fig:finit} we show the dependence of the Hamming distance $D$
on the system size. As expected from the analogy with usual critical
phenomena the spreading transition becomes sharper with increasing
system
size. Figure \ref{fig:finit} also shows that a system size of $L=20$
already gives reasonable results for the determination of the
spreading
temperature of the Coulomb glass model, provided the disorder
strength is
comparatively small.

\section{Conclusions and Outlook}
\label{sec:concl}
We have used the damage-spreading technique to examine the
low-temperature
dynamics of disordered electronic systems with localized states based
on
the Coulomb glass model.
We have found that the dynamics of the system freezes below a
spreading
temperature $T_{\rm s}$. For zero disorder this damage spreading
transition
coincides with the equilibrium phase transition within our accuracy.
At finite disorder strength, when there is no equilibrium
phase transition, the spreading point $T_{\rm s}$ is shifted to lower
temperatures. However, $T_{\rm s}$ remains finite even for larger
disorder
strengths. Consequently, there is a low temperature "phase" of the
Coulomb
glass with frozen dynamics and a high temperature phase where the
damage
spreads through the system.
In the case of finite disorder $W_0$ the spreading transition is a
purely dynamic transition which does not possess an equilibrium
counterpart. A more detailed investigation of
this transition is in progress. It is, however, hampered by
finite-size effects since the long-range interaction
severely restricts the possible system sizes in our simulations.
These limited system sizes are also the reason why the spreading
point $T_{\rm s}$ for high values of disorder could not yet be
determined
exactly.

For small disorder strengths the spreading point $T_{\rm s}$ is still
close
to the (second-order) equilibrium phase transition temperature
$T_{\rm c}$
of the system without disorder.
Since physical quantities in the vicinity of a critical point can
usually be described by scaling laws we expect the Hamming distance
$D$
to obey the homogeneity relation
\begin{equation}
  D(W_0,T)=t^\beta f\left(\frac{W_0}{t^{\varphi}}\right),\qquad
t=|T-T_{\rm c}|
\end{equation}
with the critical exponents $\varphi$ and $\beta$.
The confirmation of this scaling law and the determination of the
exponents remain a task for the future.

One might also ask, how the results change if more sophisticated
dynamical algorithms are used, that represent the physical
processes in disordered insulators better than the simple Metropolis
algorithm
with single-particle exchange with a reservoir. The question is of
particular
importance, since the properties of damage spreading
depend on the type of dynamics used in the simulation more strongly
than the thermodynamic quantities. We have begun to study the Coulomb
glass
model with distance-dependent tunneling probabilities between the
sites.
Results of this numerically much more involved investigation will be
published elsewhere.

This work was supported in part by the DAAD, by the DFG under grant
number Vo 659/1-1 and SFB 393 and by the NSF under grant number
DMR-95-10185.

\end{document}